# Nonclassical light in coupled optical systems: anomalous power distribution, Fock space dynamics and supersymmetry

R. El-Ganainy[1,2], A. Eisfeld[1] and D.N Christodoulides[3]

[1]*Max Planck Institute for the Physics of Complex systems, Nothnitzer Street 38, Dresden, Germany*
[2]*Department of Physics, Michigan Technological University, Houghton, Michigan 49931*
[3]*College of Optics & Photonics-CREOL, University of Central Florida, Orlando, Florida*

**Abstract**

We investigate the dynamics of nonclassical states of light in coupled optical structures and we demonstrate a number of intriguing features associated with such arrangements. By diagonalizing the system's Hamiltonian, we show that these geometries can support eigenstates having anomalous optical intensity distribution with no classical counterpart. These features may provide new avenues towards manipulating light flow at the quantum level. By projecting the Hamiltonian operator on Hilbert subspaces spanning different numbers of photon excitations, we demonstrate that processes such as coherent transport, state localization and Bloch oscillations can take place in Fock space. Furthermore, we show that Hamiltonian representations of Fock space manifolds differing by one photon obey a discrete supersymmetry relation.

# 1. Introduction

Investigating light behavior at the nanoscale and quantum level has been a subject of intense study in the last few decades. Such research activities have greatly benefited from recent technological developments in fabricating nano-devices and an unprecedented progress in controlling experimental conditions and measurements.

Along these lines, considerable effort has been dedicated to engineering the flow of classical light using photonic band gap materials (PBG) [1,2] and plasmonic platforms [3] with possible applications in nanoscale lasers [4], optical interconnects [2], solar cells [5] and biosensing [6] to mention a few.

On the quantum front, novel experimental techniques have allowed creation and manipulation light at few photon levels [7] as well as controlling atoms and molecules with light [8]. In addition, high-precision fabrication of quantum dots confined inside high quality microcavities has enabled the observation of many intriguing phenomena such as that of Purcell effect and vacuum Rabi splitting [9].

The synergy between nanoscale photonics and quantum optics promises even more fascinating applications. For instance, exciton dressing by a photonic band edge inside quantum wells sandwiched between heterostructure photonic crystals have been theoretically proposed [10] and plasmonic lasers (known also as spacers) have been experimentally pursued [11].

Recently, integrated waveguide systems have been proposed as a fertile platform to investigate quantum properties of light. In this context, oscillations between bunching/antibunching statistics have been demonstrated in arrays of optically coupled waveguides [12,13]. Optical lattices have been also used to investigate quantum random walks [14]. Furthermore, the effect of Anderson localization on entanglement dynamics between two photons propagating inside random waveguide arrays has been investigated both theoretically and experimentally [13, 15, 16].

In this letter we investigate dynamics of non-classical states of light inside optically coupled systems. In particular we confine our treatment to two coupled structures. We note that these schemes have been previously investigated and the system eigenmodes were used to analyze the photon statistics for number states as well as squeezed states inputs [16]. Here we focus on some of the previously overlooked features of these configurations and we demonstrate a host of novel effects.

We first consider the fundamental question of optical energy distribution in different eigenstates of nonclassical light inside such structures. In stark contrast with optical modes obtained from Maxwell's equations, we show that, at the quantum level, these systems support stationary eigenmodes having anomalous optical power profiles.

Next we consider Hilbert space sectors that span different photon number excitations, and we demonstrate the possibility of coherent transport of number states inside Fock space. Single state localization/revival effects and Bloch oscillations in configuration space are also discussed.

Finally we demonstrate an interesting discrete supersymmetry (SUSY) relation between certain Fock space manifolds. SUSY has its roots in high energy physics and studies of the standard model [17]. The concept was later extended to quantum mechanics [18] and has since then served as a powerful mathematical tool for solving otherwise formidable problems [19]. Recently, SUSY been suggested in optics as a means to engineer eigenvalue spectra of photonic systems [20]. Here we show that matrix representations of the system's Hamiltonian in Fock space sectors $S_{N+1}$ and $S_N$ (spanning all $N+1$ and $N$ photon states, respectively) form supersymmetric pairs.

## 2. Optical power distribution of nonclassical photonic eigenstates within coupled systems

Here we invsetigate the distribution of optical power associated with stationary quantum states of light inside coupled optical strcutres.

A schematic of coupled optical waveguides and cavities investigated in this work is depicted in Figs.1. (a) and (b). We start our our analysis with a brief review of classical optical modes supported by these structures.

Classically, within the coupled mode formalism [21], both systems shown in Figs.1 (a) and (b) are formally equivalent. Thus, without any loss of generality, we limit our discussion to coupled waveguide geometries. For weakly guiding structures, both scalar and paraxial approximations can be employed. This simplifies the expressions for coupling coefficients between the two waveguides without altering the general conclusions. Under these conditions, the optical field $\varphi(x,y,z)$ along two coupled waveguides of arbitrary cross section in the $x-y$ plane is expanded in terms of the electric field distribution $\varphi_{a,b}(x,y)$ of the unperturbed optical modes in waveguide $a$ and $b$: $\varphi(x,y,z) = a(z)\varphi_a(x,y) + b(z)\varphi_b(x,y)$, where $z$ is the propagation direction. It thus follows that wave propagation in this coupled waveguide system is described by[21]:

$$i\frac{d}{dz}\begin{bmatrix} a \\ b \end{bmatrix} = \begin{bmatrix} \beta_a & \kappa \\ \kappa & \beta_b \end{bmatrix}\begin{bmatrix} a \\ b \end{bmatrix} \qquad (1)$$

In Eq.(1), $\kappa > 0$ is the coupling coefficient [21] between waveguides having propagation constants of $\beta_{a,b}$ [21]. The eigenmodes of the above coupled equations are $V_e = [\cos(\alpha/2) \ \sin(\alpha/2)]^T$ and $V_o = [-\sin(\alpha/2) \ \cos(\alpha/2)]^T$, with $\tan(\alpha) = \frac{2\kappa}{(\beta_a - \beta_b)}$ and the superscript $T$ denote matrix transpose. The corresponding eigenvalues are $\beta_{e,o} = \frac{\beta_a + \beta_b}{2} \pm \sqrt{\left(\frac{\beta_a - \beta_b}{2}\right)^2 + \kappa^2}$. It thus follows that the optical power distribution of even/odd-like eigenmodes $V_{e,o}$ are respectively given by:

$$P_a / P_b = \cot^2(\alpha/2) \qquad (2.a)$$

$$P_a / P_b = \tan^2(\alpha/2) \qquad (2.b)$$

In the above expressions, $P_a = |a|^2$ and $P_b = |b|^2$ represent (up to a multiplicative constant) the optical power in waveguides $a, b$ respectively and the total optical power is $P_t = P_a + P_b$. Given that these constraints on the optical power profile (in addition to the phase relations between the two components of the mode) are solely imposed by Maxwell's equations, it is of fundamental importance to question their universality. Is it possible to circumvent these symmetry restrictions by using nonclassical states of light?

In what follows we investigate this question and we show that nonclassical states of light can have stationary eigenmodes that violate the previously mentioned classical distribution in Eqs.(2).

In order to do so, we consider general quantum states of light having a finite number of photons and propagating within these coupled optical waveguides. The system's Hamiltonian is [13-16]:

$$H = \hbar\beta_a \hat{a}^+\hat{a} + \hbar\beta_b \hat{b}^+\hat{b} + \hbar\kappa\left(\hat{a}^+\hat{b} + \hat{a}\hat{b}^+\right) \quad (3)$$

Here $\hat{a}^+$ and $\hat{a}$ are the creation and annihilation operators of photons in waveguide $a$ while $\hat{b}^+$ and $\hat{b}$ denote those associated with waveguide $b$. These operators obey the commutation relations $\left[\hat{a},\hat{a}^+\right] = \left[\hat{b},\hat{b}^+\right] = 1$ and $\left[\hat{b},\hat{a}^+\right] = \left[\hat{a},\hat{b}^+\right] = 0$.

Hamiltonian (3) is then diagonalized using the standard unitary transformation [16]:
$\begin{pmatrix}\hat{c}_e \\ \hat{c}_o\end{pmatrix} = \begin{pmatrix}\cos(\alpha/2) & \sin(\alpha/2) \\ -\sin(\alpha/2) & \cos(\alpha/2)\end{pmatrix}\begin{pmatrix}\hat{a} \\ \hat{b}\end{pmatrix}$ and we finally arrive at:

$$H = \hbar\beta_e \hat{c}_e^+\hat{c}_e + \hbar\beta_o \hat{c}_o^+\hat{c}_o \quad, \quad (4)$$

where $\beta_{e,o}$ are the same as before while $\hat{c}_{e,o}^+$ and $\hat{c}_{e,o}$ are the creation and annihilation operators of the even/odd-like eigenstates $V_{e,o}$ of equation (1) and they obey $\left[\hat{c}_{e,o},\hat{c}_{e,o}^+\right] = 1$ and $\left[\hat{c}_{e,o},\hat{c}_{o,e}^+\right] = 0$.

We now consider a state having $2N$ light quanta $|\psi_{2N,n}\rangle$ with $N \mp n$ occupying the even/odd mode, respectively: $|\psi_{2N,n}\rangle = \dfrac{\left(\hat{c}_e^+\right)^{N-n}\left(\hat{c}_o^+\right)^{N+n}}{\sqrt{(N-n)!(N+n)!}}|vac\rangle = |N-n, N+n\rangle_c$. In the above expression $|n| \leq N$ and the subscript $c$ indicates that these states are in the

even/odd basis. States with odd photon numbers can be also equally treated. It is straightforward to show that $|\psi_{2N,n}\rangle$ is an eigenstate or the Hamiltonian operator (4) with a corresponding eigenvalue $\lambda_{2N,n} = \hbar\{N(\beta_o + \beta_e) + n(\beta_o - \beta_e)\}$. Consequently, inside the Hilbert space manifold spanning $2N$ photon states, eigenvalues associated with different eigenkets $|\psi_{2N,n}\rangle$ are equidistant with a separation of $\hbar(\beta_o - \beta_e)$. Now let us investigate the average intensity in each waveguide for any such $|\psi_{2N,n}\rangle$ eigenstate. For any such a state, the statistical distribution of the optical power $P_{2N,n}^{a,b}$ is proportional to the total photon number inside each waveguide $P_{2N,n}^a \propto \langle \psi_{2N,n} | \hat{a}^+ \hat{a} | \psi_{2N,n} \rangle$ and $P_{2N,n}^b \propto \langle \psi_{2N,n} | \hat{b}^+ \hat{b} | \psi_{2N,n} \rangle$. After some simple algebra and using the relation $P_{2N,n}^t \equiv (P_{2N,n}^a + P_{2N,n}^b) \propto 2N$ for the total power, we obtain:

$$P_{2N,n}^a / P_{2N,n}^t = \frac{1}{2} - \left(\frac{\cos(\alpha)}{2N}\right) n \qquad (5.a)$$

$$P_{2N,n}^b / P_{2N,n}^t = \frac{1}{2} + \left(\frac{\cos(\alpha)}{2N}\right) n \qquad (5.b)$$

We thus arrive at a simple yet counterintuitive result: optical power distribution of nonclassical eigenstates varies linearly with the mode index $n$ and is in general different from its classical counterpart obtained by solving Maxwell's equations. We stress that these anomalous power profiles are obtained for stationary modes and their statistical distribution does not change with propagation distance.

Note that the classical power profiles are recovered for the two extreme cases when all photons occupy only one eigenmode, i.e. $n = \pm N$. We emphasize that these results hold true for any physical system modeled by the Hamiltonian (1) and are not pertinent to coupled waveguide systems. Examples of such systems are lossless coupled optical cavities and bosonic atoms in a mismatched double potential well when the inter-atomic scattering is negligible.

Given the current technical difficulties associated with preparing large photon number quantum states, it is important to note that two-photon states are sufficient for experimental demonstration of our results. In particular, optical power distribution of the two-photon state $|\psi_{2,0}\rangle = |1,1\rangle_c$ (with one photon in each mode) is $P^a_{2,0}/P^t_{2,0} = P^b_{2,0}/P^t_{2,0} = 0.5$ and deviation from classical power splitting ratios when $\beta_a \neq \beta_b$ is thus evident.

We note that directional couplers and coupled optical cavities are used as building blocks in many optical devices and it would of fundamental importance to investigate the operation of these optical components under nonclassical light excitations. This may in turn lead to altogether novel functionality and behavior.

## 3. Quantum transport and state localization in Fock space

So far we have investigated the properties of stationary quantum states of light inside coupled optical geometries. In this section we examine their dynamical properties. In particular we show that coherent quantum transport and state localization are possible in these configurations. We start by considering the Hamiltonian $H$ of Eq.(3). Any general state having $2N$ photons can be written as a superposition of the kets $|n_{2N}\rangle = \frac{(\hat{a}^+)^{N-n}(\hat{b}^+)^{N+n}}{(N-n)!(N+n)!}|vac\rangle = |N-n, N+n\rangle$, with $N \mp n$ photons in waveguides $a$ and $b$ respectively. Note that the set $|N-n, N+n\rangle$ is different from previously embloyed bases of $|N-n, N+n\rangle_c$. The above choice of $|n_{2N}\rangle$ automatically emphasizes the inversion symmetry around state $|0_{2N}\rangle = |N, N\rangle$. States with odd photon numbers can be easily treated using a different set of bases.

The spectral representation of the Hamiltonian $H$ inside any such $2N$ photon manifold $S_{2N}$ then reads:

$$H_{2N} = 2\hbar\beta_{avg}N\hat{I} + \hbar\Delta\sum_{n=-N}^{N}n|n_{2N}\rangle\langle n_{2N}| + \\ \hbar\kappa\sum_{n=-N}^{N-1}\sqrt{(N+n+1)(N-n)}\left(|n_{2N}\rangle\langle(n+1)_{2N}| + |(n+1)_{2N}\rangle\langle n_{2N}|\right) \quad (6)$$

Where $\Delta = \beta_b - \beta_a$, $\beta_{avg} = (\beta_a + \beta_b)/2$ and $\hat{I}$ is the identity operator. The unitary evolution of any arbitrary wavefunction inside this subspace is given by $|\varphi(z)\rangle = \sum_{n=-N}^{N}\exp(-2iN\beta_{avg}z)\chi_n(z)|n_{2N}\rangle$ where the dynamics of $\chi_n(z)$ with a set of initial conditions $\chi_n(0)$ follows directly from $H_{2N}$ in Eq.(6):

$$i\frac{d\chi_{-N}}{dz} = -\Delta N\chi_{-N} + \kappa\sqrt{2N}\chi_{-N+1} \\ i\frac{d\chi_n}{dz} = \Delta n\chi_n + \kappa(g_n\chi_{n-1} + g_{n+1}\chi_{n+1}) \quad , \quad -N < n < N \quad (7) \\ i\frac{d\chi_N}{dz} = \Delta N\chi_N + \kappa\sqrt{2N}\chi_{N-1}$$

In Eq.(7) $g_n = \sqrt{(N+n)(N-n+1)}$ and thus the coupling coefficients are symmetric around $n = 0$. For matched waveguides with $\Delta = 0$, equation (7) is mathematically equivalent to a spin lattice model envisioned for coherent quantum transport [22] and later extended to optical arrays [23]. In contrast to similar behavior in real lattices [22,23], we emphasize that, in this work, the aforementioned processes occur in Fock space.

Figure 2(a) shows a schematic representation to the Fock space array described by Eq.(7). Coupling coefficients corresponding to 20-photon manifold for waveguides normalized parameters of $\kappa = 0.1$ and $\Delta = 0$ are plotted in Fig.2(b). As shown in Fig.2(c), a complete transport of an initial state $|-6_{20}\rangle = |16,4\rangle$ into its mirror symmetry (with respect to $n = 0$) state $|6_{20}\rangle = |4,16\rangle$ occurs after a normalized (with respect to $\max(g_n) \times \kappa$) propagation distance of approximately $16.5$. Contrary to similar effects occurring in carefully engineered real space lattice systems [23], Fock space quantum transport predicted here naturally arises as a result of the bosonic statistics of photons.

Next we consider mismatched waveguides, i.e. $\Delta \neq 0$. From diagonalizing the Hamiltonian (3), we have found that the eigenvalues are equidistant with a spacing $\hbar(\beta_o - \beta_e)$. Consequently, the Fock arrays described by (7) closely resemble Bloch arrays having Wannier-Stark eigenmodes [24] and are thus expected to exhibit state revivals (diffraction is followed by refocusing) and Bloch oscillations [25,26].

Before we illustrate our results, we again emphasize that in this work, the revivals effects and Bloch oscillations are a direct outcome of the photon statistics and they take place in Fock space as opposed to their counterpart real space processes.

Fig.3.(a) depicts three (5[th], 10[th], and 15[th] from left to right) of these localized stationary solutions of Eq.(7) when $\Delta = 1$, $\kappa = 0.1$ and the total number of photons is $20$. In contrast to an ideal Bloch array where all Wannier-Stark modes are shifted replicas of the same mode profile [25,26], here the eigenvectors are not identical. Figure 3(b) demonstrates dynamical evolution for the initial state $\left|0_{10}\right\rangle=\left|10,10\right\rangle$ where localization and revival effects are evident. Figure 3(c) depicts surface Bloch oscillations when the initial wavepacket has the shape of a truncated Gaussian profile centered at the state $\left|-10_{20}\right\rangle=\left|20,0\right\rangle$, where perfect oscillations can be observed. The dynamics of the same initial state inside a Bloch waveguide array with identical coupling constants are shown in Fig.3(d). In this latter case we observe localization without perfect oscillations. This discrepancy is due to the eigenvalue distribution in both systems. Bloch arrays exhibit equidistant eigenvalues only for infinite lattice and any edge effect can disturb this uniformity and destroy oscillations. On the other hand, the array described by (7) has uniform eigenvalue distribution irrespective of the number of elements (or equivalently the number of photons) and complete Bloch oscillations are thus preserved. Similar effects have been reported using classical light propagating in Glauber-Fock arrays

## 4. Discrete SUSY connection between Fock space sub-manifolds

In this section we show that matrix representations of the system's Hamiltonian in Fock space sectors $S_{N+1}$ and $S_N$ (spanning all $N+1\ \&\ N$ photon states, respectively) form supersymmetric pairs. It is important to note that, in this work SUSY emerges naturally and is not imposed on the system.

We employ the bases $|n_{N+1,n}\rangle_g = |N+1-n, n\rangle$, $n = 0, 1, ... N+1$, and we denote the projection of $H$ on $S_{N+1}$ by $H_{N+1}$ and its tridiagonal matrix representation in terms of $|n_{N+1,n}\rangle_g$ by $\mathcal{H}_{N+1}$. Note that $|n_{N+1}\rangle$ span $S_{N+1}$ for every $N$, i.e. any general states with even or odd photon numbers can be represented by $|n_{N+1}\rangle_g$ and hence the subscript $g$. By noting that the eigenfunctions of $H_{N+1}$ are given by $|\psi_{N+1,n}\rangle_g = |N+1-n, n\rangle_c$ and by recalling that $\beta_e \leq \beta_o$, we find that $|\psi_{N+1,0}\rangle_g = |N+1, 0\rangle_c$ is the $S_{N+1}$ ground state with corresponding eigenvalue of $\hbar(N+1)\beta_e$. Consequently, the ground state of the shifted Hamiltonian $\tilde{\mathcal{H}}_{N+1} \equiv \mathcal{H}_{N+1} - \hbar(N+1)\beta_e I_{N+1}$, with $I_{N+1}$ being an identity matrix with dimensions $N+1$, is zero. Using Cholesky factorization, we obtain $\tilde{\mathcal{H}}_{N+1} = LL^T$, where $L$ is a lower diagonal matrix and the superscript $T$ denote matrix transpose. We now investigate the properties of the matrix $L^T L$. It is easy to show that both $L^T L$ and $\tilde{\mathcal{H}}_{N+1}$ exhibit identical spectra [20] and that their eigenfunctions are related by $L^T |\varphi_k^{(1)}\rangle = |\varphi_k^{(2)}\rangle$ and $L|\varphi_k^{(2)}\rangle = |\varphi_k^{(1)}\rangle$.

Here in contrast to SUSY quantum mechanics [19], the zero ground state of $L^T L$ does not disappear. However we now show that under certain condition, this eigenstate can be isolated from the rest of the spectrum. Since $L^T L$ has a ground state with eigenvalue zero, it follows that the determinant $|L^T L| = 0$. Using $|L^T L| = |L^T||L|$ and by noting that $|L^T| = |L|$, it thus follows that $|L^T| = 0$. Consequently, the matrix $L^T$ also has a zero eigenvector which we denote by $|u\rangle$ and it thus follows that $|u\rangle$ is also the ground state of

$\tilde{\mathcal{H}}_{N+1}$ and is given by $|u\rangle = |N+1, 0\rangle_c$. By invoking the explicit forms of

$$|u\rangle = [u_1 \ u_2 \ ... \ u_{N+1}] \text{ and } L^T = \begin{bmatrix} l_{1,1} & l_{1,2} & ... & l_{1,N+1} \\ 0 & l_{2,2} & ... & l_{2,N+1} \\ ... & & ... & ... \\ 0 & 0 & ... & l_{N+1,N+1} \end{bmatrix}, \text{ and by using } L^T |u\rangle = 0 \text{ we}$$

obtain $l_{N+1,N+1} u_{N+1} = 0$. Since $u_{N+1} \neq 0$, we find that $l_{N+1,N+1} = 0$ and $L^T L = \begin{bmatrix} A & 0 \\ 0 & 0 \end{bmatrix}$, where $A$ is a symmetric tridiagonal $N \times N$ matrix that contains the rest of the spectrum of $L^T L$ excluding the null eigenvalue. Thus both $\tilde{\mathcal{H}}_{N+1}$ and $A$ are connected through discrete supersymmetry [20]. It is also straightforward to check that both Hermitian matrices $A$ and $\tilde{\mathcal{H}}_N \equiv \mathcal{H}_N - \hbar N \beta_e I_N + \hbar(\beta_o - \beta_e) I_N$ must be similar. In the above used bases, the two matrices turn out to be identical and $L^T L = \begin{bmatrix} \tilde{\mathcal{H}}_N & 0 \\ 0 & 0 \end{bmatrix}$. The matrices $\tilde{\mathcal{H}}_{N+1}$ and $\tilde{\mathcal{H}}_N$ are thus supersymmetric and the eigenvectors $|\varphi_N^n\rangle$ of $\mathcal{H}_N$ can be calculated from those of $\mathcal{H}_{N+1}$ by applying $L^T |\varphi_{N+1}^{n+1}\rangle$ and then discarding the last zero element. Figure 4 summarizes these relations between the spectrum and eigenvectors of $\tilde{\mathcal{H}}_{N+1}$ and $\tilde{\mathcal{H}}_N$. This property is a direct consequence of the photon statistics and will thus prevail for any bosonic system in two coupled wells when the inter-body interaction is negligible.

Finally we remark that $\mathcal{H}_N$ can be represented in terms of a spin $N/2$ particle [22] and the standard interpretation of SUSY as a connection between two particles with $1/2$ spin difference is then recovered.

## 5. Conclusions

We have investigated the problem of nonclassical light transport in coupled optical arrangements and we have demonstrated a number of intriguing and previously overlooked features in these systems. We have shown that these structures can support quantum states of light that exhibit anomalous optical power distribution and having no counterpart whatsoever in classical optics. Fock space quantum transport and state localization were also predicted and we have demonstrated the possibility of surface Bloch oscillations and have shown that perfect revivals can still be maintained despite edge effects. Finally we have shown that Fock space manifolds differing by one photon obey discrete supersymmetry relations. The aforementioned processes are not a result of careful engineering of the system's parameters but instead arise naturally in coupled optical configurations due to the bosonic nature of photon statistics.

# References

1. S. John, "Strong localization of photons in certain disordered dielectric superlattices", Phys Rev. Lett. **58,** 2486 (1987)**,** E. Yablonovitch,"Inhibited Spontaneous Emission in Solid-State Physics and Electronics" Phys. Rev. Lett. **58,** 2059 (1987)

2. J. Joannopoulos, S.G. Johnson, J.N. Winn and R. D. Meade**,** *Photonic crystals, molding the flow of light*, Princeton University Press (2008).

3. S. A. Maier, *Plasmonics: Fundamentals and Applications*, Springer (2010).

4. D.J. Bergman and M.I. Stockman, "Surface Plasmon Amplification by Stimulated Emission of Radiation: Quantum Generation of Coherent Surface Plasmons in Nanosystems" Phys Rev. Lett. **90**, 027402 (2003)

5. E. Garnett and P. Yang, "Light Trapping in Silicon Nanowire Solar Cells" *Nano Lett.* **10**, 1082, (2010).

6. P.N. Prasad, *Introduction to Biophotonics*, Wiley-Interscience, (2003)

7. P.G. Kwiat, K. Mattle, H. Weinfurter, and A. Zeilinger, "New High-Intensity Source of Polarization-Entangled Photon Pairs" Phys. Rev. Lett. **75**, 4337 (1995).

8. S. Chu, J.E. Bjorkholm, A. Ashkin and A. Cable, "Experimental Observation of Optically Trapped Atoms" Phys. Rev. Lett. **57**, 314 (1986).

9. G. Khitrova, H. M. Gibbs, M. Kira, S. W. Koch and A. Scherer, "Vacuum Rabi splitting in semiconductors", Nature Physics 2, **81** (2006).

10. S. Yang and S. John, "Exciton dressing and capture by a photonic band edge", Phys. Rev. B **75**, 235332 (2007).

11. M. A. Noginov, G. Zhu, A. M. Belgrave, R. Bakker, V. M. Shalaev, E. E. Narimanov, S. Stout, E. Herz, T. Suteewong and U. Wiesner, "Demonstration of a spaser-based nanolaser" Nature**, 460**, 1110 (2009).

12. Y. Bromberg, Y. Lahini, and Y. Silberberg, "Bloch Oscillations of Path-Entangled Photons" Phys. Rev. Lett. **105**, 263604 (2010)

13. Y. Lahini, Y. Bromberg, Y. Shechtman, A. Szameit, D. N. Christodoulides, R. Morandotti, and Y. Silberberg, "Hanbury Brown and Twiss correlations of Anderson localized waves" Phys. Rev. A **84**, 041806(R) (2011)

14. A. Peruzzo et al, "Quantum Walks of Correlated Photons" *Science,* **329**, 1500 (2010)

15. A.F. Abouraddy, G.D. Giuseppe, D.N. Christodoulides, and B.E.A. Saleh, "Anderson localization and colocalization of spatially entangled photons" Phys. Rev. A **86**, 040302(R) (2012)**;** G. Di Giuseppe, L. Martin, A. Perez-Leija, R. Keil, F. Dreisow, S. Nolte, A. Szameit, A. F. Abouraddy, D. N. Christodoulides, and B. E. A. Saleh "Einstein-Podolsky-Rosen Spatial Entanglement in Ordered and Anderson Photonic Lattices" Phys. Rev. Lett. **110**, 150503 (2013)

16. W.K. Lai, V. Buek, and P.L. Knight, "Nonclassical fields in a linear directional coupler" Phys. Rev. A **43**, 6323 (1991)

17. J. Wess and B. Zumino, Nucl. Phys. B **70**, 39 (1974).

18. E. Witten, Nucl. Phys. B **185**, 513 (1981).

19. F. Cooper, *Supersymmetry in Quantum Mechanics***,** World Scientific Pub Co Inc ( 2002)

20. M.A. Miri, M. Heinrich, R. El-Ganainy, and D. N. Christodoulides, "Supersymmetric optical structures", submitted.

21. K. Okamoto, *Fundamentals of Optical Waveguides*, Academic Press (2005).

22. M. Christandl, N. Datta, A. Ekert, and A. J. Landahl, Phys. Rev. Lett. **92**, 187902 (2004).

23. A. Perez-Leija, R. Keil, A. Kay, H. Moya-Cessa, S. Nolte, L. C. Kwek, B. Rodrıguez-Lara, A. Szameit, and D. N. Christodoulides, "Coherent quantum transport in photonic lattices" Phys. Rev. A **87**, 012309 (2013).

24. C. Zener, "A theory of the electrical breakdown of solid dielectrics," Proc. R. Soc. London A **145**, 523 (1934).

25. U. Peschel, T. Pertsch, and F. Lederer, "Optical Bloch oscillations in waveguide arrays," Opt. Lett. **23**, 1701 (1998)

26. T. Pertsch, P. Dannberg, W. Elflein, and A. Bräuer, "Optical Bloch oscillations in temperature tuned waveguide arrays," Phys. Rev. Lett. **83**, 4752 (1999).

27. R. Keil, A. Perez-Leija, P. Aleahmad, H. Moya-Cessa, S. Nolte, D. N. Christodoulides, and A. Szameit "Observation of Bloch-like revivals in semi-infinite Glauber–Fock photonic lattices" Optics Letters **37**, 3801 (2012)

# Figure captions

Fig.1. (Color online) Schematics of coupled optical systems under consideration are depicted in (a) and (b). In (a) quantum states of light propagate inside a coupled mismatched waveguide system while in (b) photons transfer back and forth between two coupled optical cavities. Within coupled mode theory, the unitary evolution of the systems (a) and (b) is equivalent to tunneling dynamics of bosonic particles within a double potential well when inter-particle scattering is negligible (c).

Fig.2. (Color online) Schematic visualization of Eq. (6) in terms of coupled arrays. (b) Left and right coupling profile of Eq.(7) for a 20 photons propagating inside a zero detuned directional coupler having a normalized coupling of $\kappa = 0.1$. (c) Demonstration of transport effect. The state $|-6_{20}\rangle = |16,4\rangle$ is transported into $|6_{20}\rangle = |4,16\rangle$ after a normalized propagation distance of $z \cong 16.5$.

Fig.3. (Color online) (a) three different localized solutions for quantum states of light inside mismatched directional coupler having the normalized parameters $\Delta = 1$, and $\kappa = 0.1$. (b) Revival effects for an input state $|0_{20}\rangle = |10,10\rangle$. Note that these effects occur in Fock space. (c) Fock space surface Bloch oscillations of an initial state having a Gaussian profile for physical coupling parameters. (d) Evolution of the initial state in (c) if the coupling coefficients were uniform as typically used in waveguide arrays.

Fig.4. (Color online) SUSY symmetry between the shifted Hamiltonians $\tilde{\mathcal{H}}_{N+1}$ and $\tilde{\mathcal{H}}_{N}$. Apart from the ground state of $\tilde{\mathcal{H}}_{N+1}$, both matrices share the same spectrum. Their eigenvectors can be exchangeably generated through application of operators $L$ or $L^T$ on the corresponding space. Note that $L^T|\varphi_{N+1}^{n+1}\rangle$ must be followed by removing the last element of the resulting vector while $L|\varphi_N^n\rangle$ should be preceded by appending a zero element to $|\varphi_N^n\rangle$.

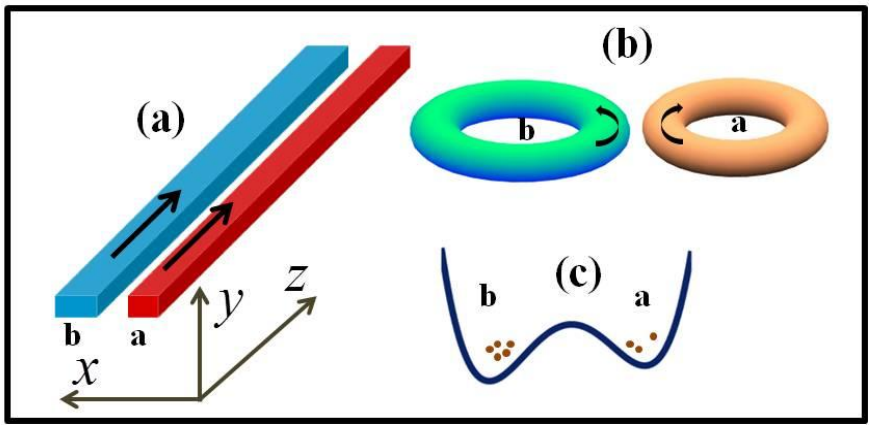

Fig.1

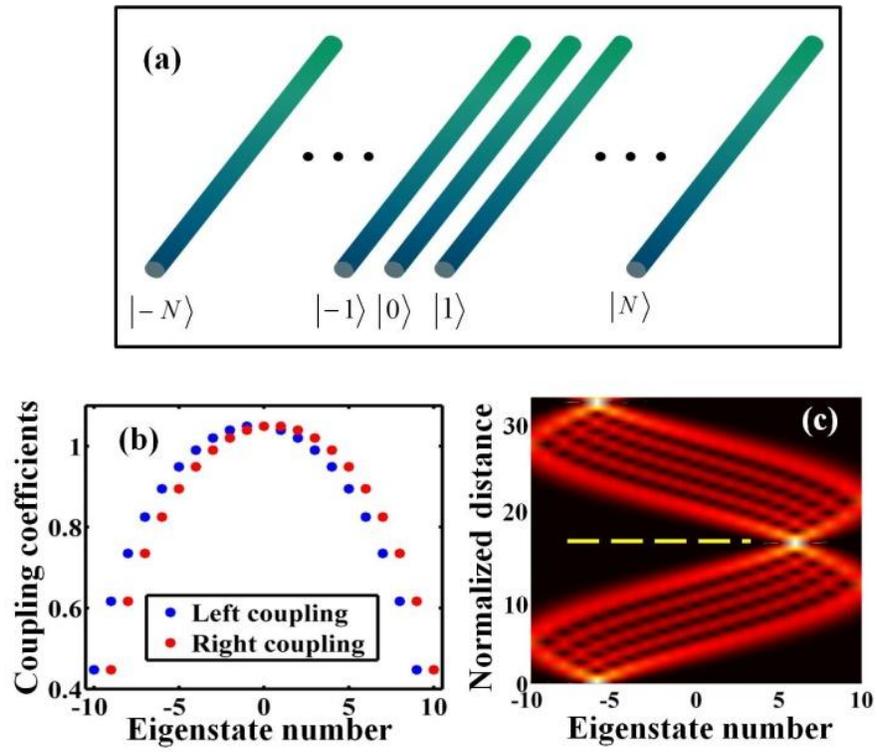

Fig.2

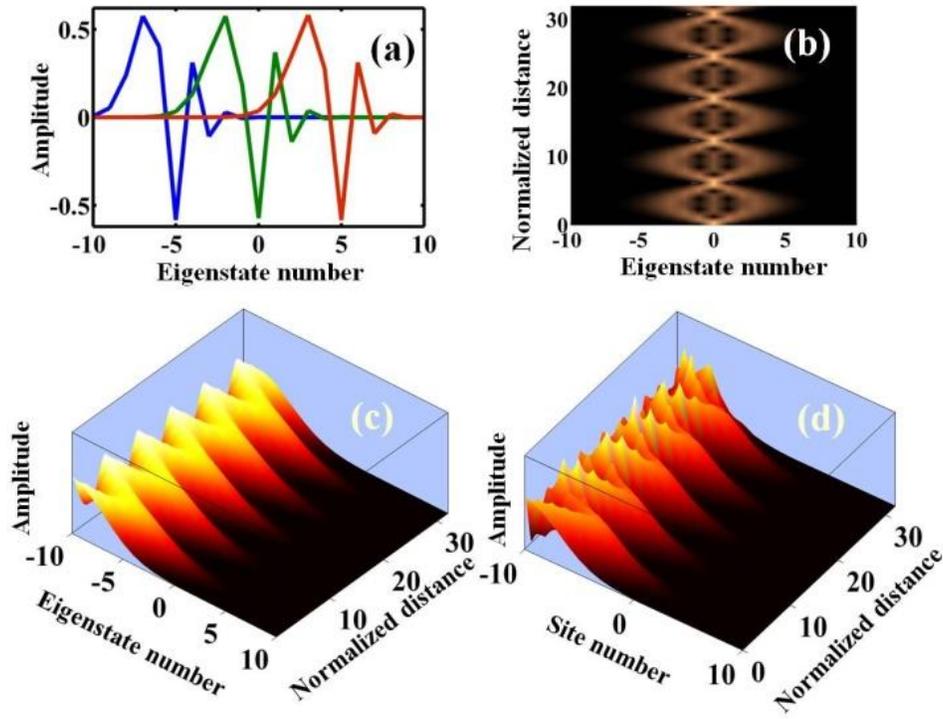

Fig.3

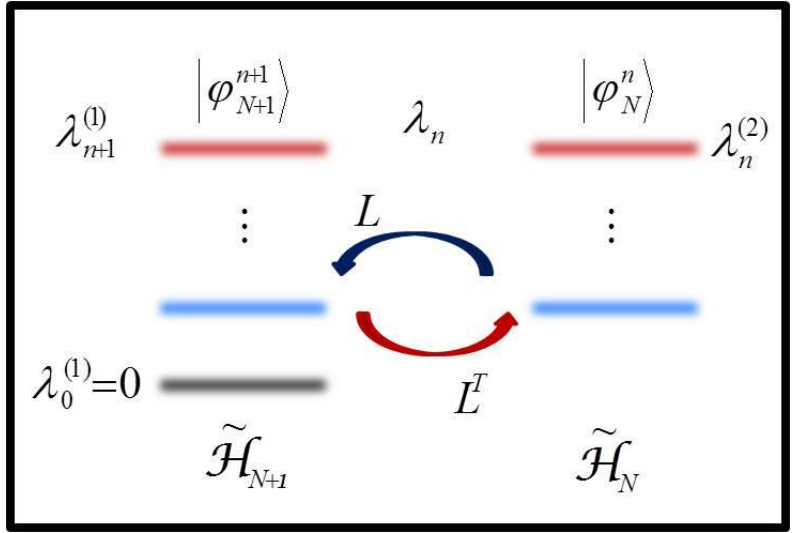

Fig.4